\documentclass[aps,twocolumn,showpacs,prl,superscriptaddress,unsortedaddress]{revtex4}
\usepackage{epsf}
\usepackage{graphicx}
\newcommand{\etal}{{\it et al.}}

\begin{document}

\title{Anomalous dispersion in the autocorrelation of ARPES data of Bi$_2$Sr$_2$CaCu$_2$O$_{8+\delta}$}
         
\author{U. Chatterjee}
\affiliation{Department of Physics, University of Illinois at Chicago, Chicago, IL 60607}
\affiliation{Materials Science Division, Argonne National Laboratory, Argonne, IL 60439}
\author{M. Shi}
\affiliation{Department of Physics, University of Illinois at Chicago, Chicago, IL 60607}
\affiliation{Swiss Light Source, Paul Scherrer Institut, CH-5232 Villigen, Switzerland}
\author{A. Kaminski}
\affiliation{Ames Laboratory and Department of Physics and Astronomy, Iowa State University, 
Ames, IA 50011}
\author{A. Kanigel}
\affiliation{Department of Physics, University of Illinois at Chicago, Chicago, IL 60607}
\author{H. M. Fretwell}
\affiliation{Ames Laboratory and Department of Physics and Astronomy, Iowa State University, 
Ames, IA 50011}
\author{K. Terashima}
\author{T. Takahashi}
\affiliation{Department of Physics, Tohoku University, 980-8578 Sendai, Japan}
\author{S. Rosenkranz}
\affiliation{Materials Science Division, Argonne National Laboratory, Argonne, IL 60439}
\author{Z. Z. Li}
\author{H. Raffy}
\author{A. Santander-Syro}
\affiliation{Laboratorie de Physique des Solides, Universite Paris-Sud, 91405 Orsay Cedex, France}
\author{K. Kadowaki}
\affiliation{Institute of Materials Science, University of Tsukuba, Ibaraki 305-3573, Japan}
\author{M. Randeria}
\affiliation{Department of Physics, Ohio State University, Columbus, OH  43210}
\author{M. R. Norman}
\affiliation{Materials Science Division, Argonne National Laboratory, Argonne, IL 60439}
\author{J. C. Campuzano}
\affiliation{Department of Physics, University of Illinois at Chicago, Chicago, IL 60607}
\affiliation{Materials Science Division, Argonne National Laboratory, Argonne, IL 60439}

\date{\today}

\begin{abstract}
We find that peaks in the autocorrelation of angle resolved photoemission spectroscopy data of Bi$_2$Sr$_2$CaCu$_2$O$_{8+\delta}$ in the superconducting state show dispersive behavior for binding energies smaller than the maximum superconducting energy gap.  For higher energies, though, a striking anomalous dispersion is observed that is a consequence of the interaction of the electrons with collective excitations.  In contrast, in the pseudogap phase, we only observe dispersionless behavior for the autocorrelation peaks.  The implications of our findings in regards to Fourier transformed scanning tunneling spectroscopy data are discussed.
\end{abstract}

\maketitle

A key challenge in condensed matter physics is the detection of fluctuating order \cite{RMP}. An example is  the striking checkerboard pattern observed in Fourier transformed scanning tunneling spectroscopy (FT-STS) experiments on Ca$_{2-x}$Na$_x$CuO$_2$Cl$_2$ \cite{HANAGURI}.  Related patterns seen in the cuprate superconductor Bi$_2$Sr$_2$CaCu$_2$O$_{8+\delta}$ are sensitive to the bias energy of the experiment. They have been interpreted as arising from scattering between regions of high density of states \cite{HOFFMAN,MCELROY}. We have recently shown that an autocorrelation analysis of angle resolved photoemission spectroscopy (ARPES) data, $C(\textbf q,\omega)=\sum_{\textbf k} I(\textbf{k+q},\omega) I(\textbf{k},\omega)$ - the product of measured ARPES intensities at two different momenta at fixed energy $\omega$ separated by a momentum transfer $\textbf{q}$, effectively  maps out the momentum resolved joint density of states \cite{AC_ARPES}. 
This, in turn, provides insight into what features of the electronic structure might play a dominant role in electron scattering phenomena. In particular, we found that in the superconducting state for $\omega <  \Delta$ (where $\Delta$ is the maximum superconducting gap), peaks in the autocorrelations are dispersive, as in the FT-STS data \cite{HOFFMAN,MCELROY}, while in the pseudogap phase, the peaks are non-dispersive, again as seen in FT-STS data \cite{ALI,STM_ZTPG}.  In the first case,
the dispersive peaks are associated with the ends of the constant energy contours (`bananas') which are a consequence of the d-wave anisotropy of the superconducting gap.  In the second case, the dispersionless peaks are associated with scattering from the tips of the Fermi arcs, that we found to be binding energy independent.

In this paper we show that in an energy range $\Delta < \omega < \omega_{dip}$, where $\omega_{dip}$ is the energy where a sharp minimum is observed in the ARPES spectra near the antinode \cite{PEAKDIPHUMP}, the autocorrelation peaks in the superconducting state show an anomalous dispersion, whose origin is quite different from that of the dispersion observed for $\omega < \Delta$. This behavior is in contrast to the
pseudogap phase, where the peaks remain dispersionless.
We further show that at still higher energies, $\omega > \omega_{dip}$, autocorrelations from both the pseudogap and superconducting states are non-dispersive, which is also evident in
FT-STS data \cite{STM_ZTPG}.  In both energy ranges, this behavior can be traced to the fact
that the autocorrelation peaks track the antinodal $(\pi,0)-(\pi,\pi)$ dispersion, which in the
superconducting state has an anomalous `S' shape due to the coupling of the electrons to
collective excitations.  This `S' shape is not observed in the pseudogap phase.
 
Our data are from single crystal samples of optimally doped Bi$_2$Sr$_2$CaCu$_2$O$_{8+\delta}$  (T$_c$=90K), in the superconducting state  at T=40K and in the pseudogap phase at T=140K. Experimental details are the same as those described in Ref.~\onlinecite{AC_ARPES}.  For this sample, $\Delta=45$ meV, and $\omega_{dip}=75$ meV. In the superconducting state, constant energy intensity maps have characteristic `banana' shapes, shown in Fig.~1a for $\omega$=18 meV. These bananas increase in extent with increasing binding energy due to the d-wave anisotropy of the gap. The highest intensities are observed at the tips of the bananas, these being turning points in the constant energy contours. In this energy range, antinodal regions show little intensity in the superconducting state.

The bananas lead to peaks in the autocorrelation $C(\textbf q,\omega)$ that are highly dispersive. 
Fig.~1b shows the autocorrelation map corresponding to Fig.~1a. The vectors $\bf{q_1-q_7}$ spanning the tips of the bananas correspond to those of the octet model \cite{HOFFMAN,MCELROY,WANG}. In Fig.~1a we highlight an additional vector not discussed in our earlier work \cite{AC_ARPES},  
$\textbf{q}^{\prime}$ \cite{CAPRIOTI}, previously described by McElroy \etal \cite{MCELROYAC}, arises from the high joint density of states connecting  diagonally opposite nodal regions, and is nearly dispersionless for a large range of $\omega$ due to the high Fermi velocity along the nodal direction. On the other hand, the ARPES intensity maps in the pseudogap phase, Fig.~1c, show high intensity lines due to gapless excitations known as Fermi arcs \cite{ARCS}. Unlike the superconducting bananas, these Fermi arcs do not change in extent with binding energy (note that the value of 18 meV in the superconducting state, Fig. 1a, was chosen so that the length of the banana is the same as that of the arc). This leads to non-dispersive features in the autocorrelations (Fig.~1d) in the pseudogap phase \cite{AC_ARPES}, unlike the dispersive ones seen in the superconducting state \cite{AC_ARPES,MCELROYAC}. Although all the $\textbf{q}$-space features present in the superconducting $C(\textbf q,\omega)$ also appear to be present in the pseudogap $C(\textbf q,\omega)$, a crucial difference is their dispersionless character in the
pseudogap case (the autocorrelations versus binding energy along the bond direction are shown in Fig.~2). These features are rather smeared due to  the much larger intrinsic and thermal spectral broadening in the pseudogap phase. Since the pseudogap is not a true gap but rather a suppression of low energy spectral weight,  intensity is also present in the antinodal regions of the zone. As we will see below, this has some consequence in regards to autocorrelation features in the pseudogap phase.

\begin{figure}
\includegraphics[width=3.6in]{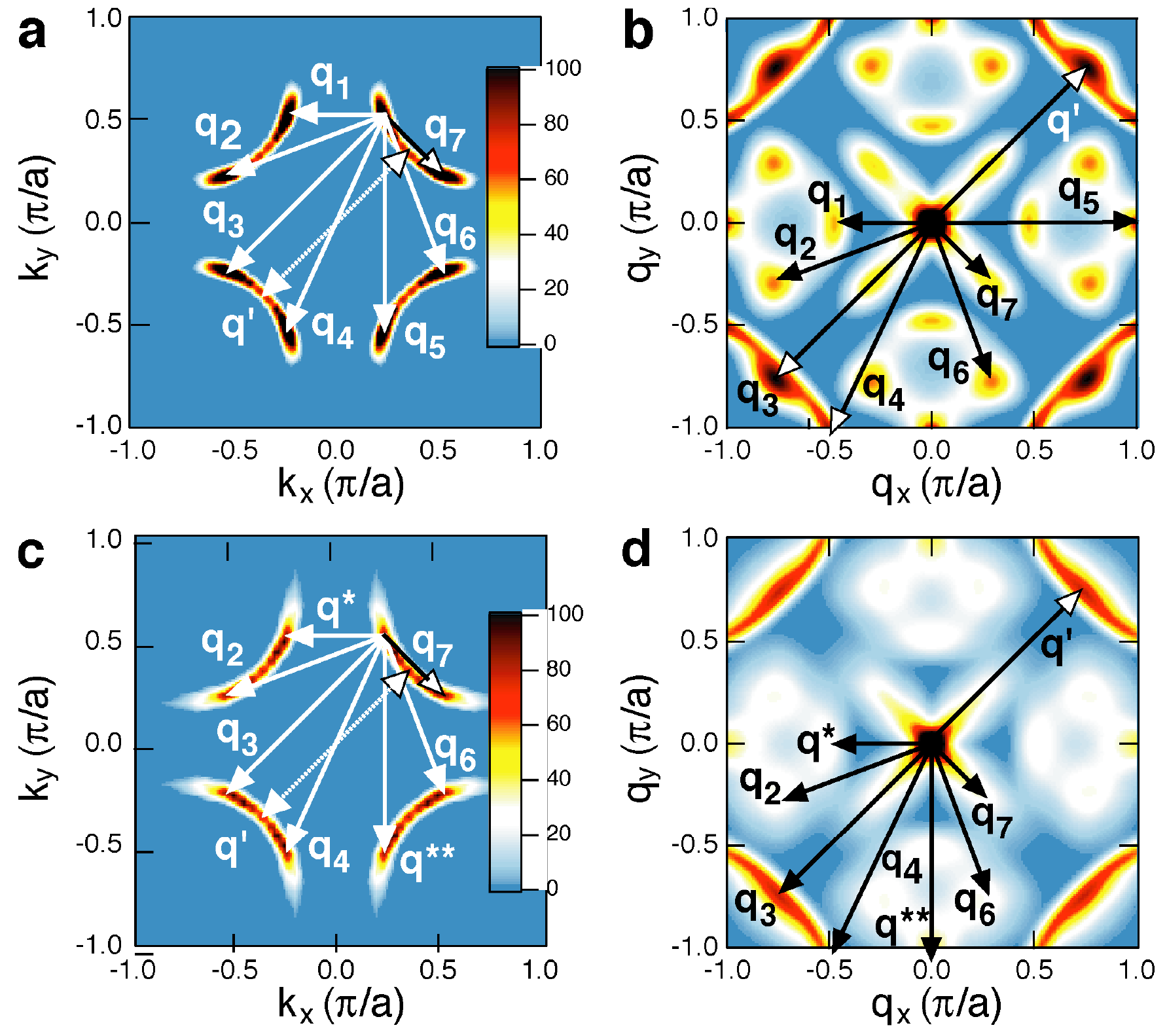}
\caption{(Color online) (a) Constant energy intensity map and (b) autocorrelation at 18 meV in the superconducting state, showing all the vectors consistent with the octet model. The intensity is non-uniform along the bananas, with high intensity points being confined to the tips of the bananas. (c) and (d) are the analogous quantities in the pseudogap phase at 0 meV.
Note that the length of the banana at 18 (meV) (Fig. 1a) is the same as that of the arc at 0 meV (Fig. 1c).}
\label{fig1.eps}
\end{figure}
 
We now discuss the most interesting energy range, for $\Delta < \omega < \omega_{dip}$. Fig.~2a shows the dispersion for the superconducting state and Fig.~2b that for the pseuodgap phase. (For the pseudogap state, we use the notation $\bf{q^*}$ and $\bf{q^{**}}$ for vectors at low energies which connect the arc tips. For higher energies, we again use the notation $\bf{q_1-q_7}$ of the octet
model.) In the pseudogap phase, the vectors $\bf{q_1}$ and $\bf{q_5}$ are non-dispersive in this energy window. On the other hand, the superconducting autocorrelation is strikingly different: $\bf{q_1}$ and $\bf{q_5}$ exhibit an anomalous dispersion,  anomalous in the sense that with increasing binding energy, $\bf{q_1}$ and $\bf{q_5}$ disperse in {\it opposite} direction to their equivalent $\bf{q_1}$ and $\bf{q_5}$ in the low energy range discussed above. 

To understand the origin of this anomaly, we plot in Fig.~3a the dispersion along $(\pi,0)-(\pi,\pi)$ from the ARPES momentum distribution curves (MDCs) (squares) and that of $\bf{q_1}$ (circles), as well as their intensities in Fig.~3b.  From Fig.~3b it is easy to see that the intensity profiles of both the MDCs and the $\bf{q_1}$ peaks follow each other as a function of energy. This is expected, as the $\bf{q_1}$ peak is related to the high joint density of states associated with the antinodal regions of the zone, where the constant energy contours are essentially parallel to one another. This same correlation is found in the dispersion of  both $\bf{q_1}$ and  the antinodal MDC as shown in Fig.~3a (for this comparison, the MDC  ${\textbf k}$ is doubled in value so as to properly correlate it with the magnitude of $\bf{q_1}$).  For $\omega > \Delta$, the MDC peak disperses towards ($\pi$,0) and after reaching a minimum,  starts to disperse away from ($\pi$,0). This dispersion continues up to the energy of the dip in the spectral function, and for still higher energies, remains almost non-dispersive up to the energy of the hump in the ARPES spectra. This same `S' shaped dispersion is also visible in the dispersion of $\bf{q_1}$ in Fig.~3a.

\begin{figure}
\includegraphics[width=3.4in]{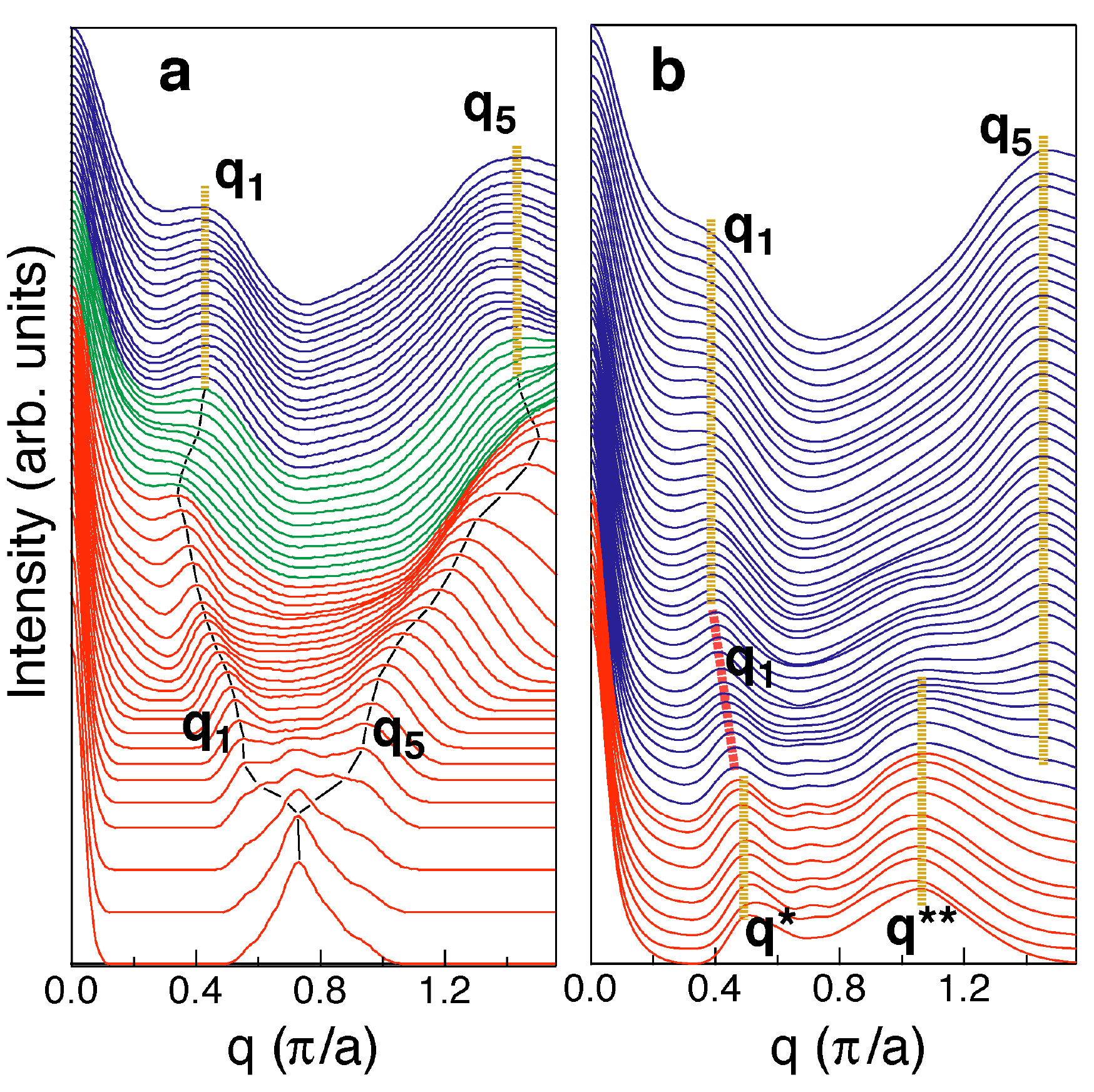}
\caption{(C0lor online) Plots of $C(\textbf q,\omega)$ along the bond direction for (a) the superconducting state, and (b) the pseudogap phase.  The bottom curve is at 0 meV and the top curve at 98 meV. Consecutive curves are separated by 2 meV.}
\label{fig2.eps}
\end{figure}

\begin{figure}
\includegraphics[width=3.4in]{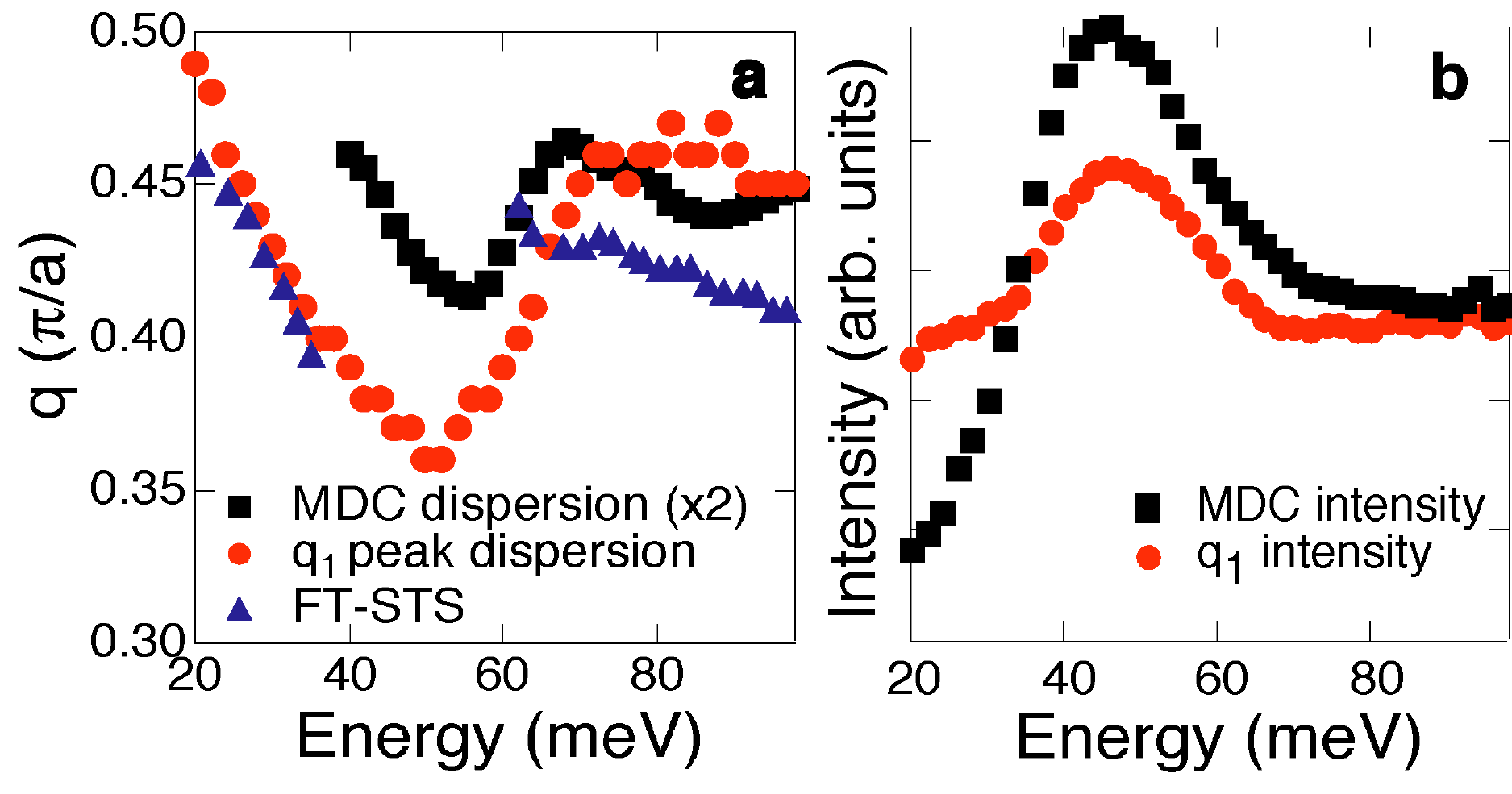}
\caption{(Color online) (a) Plot of (a) the dispersions and (b) the intensities of the MDC along $(\pi,0)-(\pi,\pi)$ (black squares) and the autocorrelation vector $\bf{q_1}$ (circles) in the energy range of 20 to 100 meV in the superconducting state.  The MDC dispersion  ${\textbf k}$ is multiplied by a factor of two so as
to compare to the autocorrelation  ${\textbf q}$. The similarities of the two point to the origin of the anomalous dispersion in the autocorrelation as discussed in the text.  For comparison, we also
plot the negative bias FT-STS results of Fig. 4c from McElroy \etal \cite{STM_ZTPG}.}
\label{fig3.eps}
\end{figure}

Normally, in the vicinity of the chemical potential, an electronic state would show a monotonic dispersion towards the chemical potential. But the situation where the dispersion turns back in a non-monotonic fashion typically arises when electrons interact with a bosonic mode \cite{MDCINGAP}. This interaction renormalizes the dispersion, leading to an `S' shape when the dispersion is traced using MDCs, and this effect is mirrored in the dispersion of the autocorrelation peaks.  It is still highly controversial whether 
the bosonic mode in question is a spin mode or a phonon.
The fact that this `S' is not seen in the pseuodgap phase would be in support of
a spin mode interpretation, as the spin excitations do change dramatically when passing through  T$_c$.  This was the original argument of Ref.~\onlinecite{ARPES_MAGNON}, but counter arguments in
favor of a phonon have been presented by others \cite{ARPES_PHONON}.
Irrespective of its origin, the clear signature of such collective excitations presents a new aspect to the features in the autocorrelation of ARPES data.
For $\omega > \omega_{dip}$, this renormalization effect is no longer evident, and one finds rather
dispersionless peaks in both the MDCs and the autocorrelations.  The practically dispersionless
nature of the MDCs along $(\pi,0)-(\pi,\pi)$ for optimal doped samples can be seen in our earlier ARPES studies \cite{BILAYER}, and is a consequence of the existence of a high energy scale associated with the `hump' in the spectral function, which moves to higher binding energy with decreasing
doping \cite{JC99}.

This high energy behavior is summarized in Fig.~4.  In the superconducting state,
antinodal high intensity regions are parallel to each other in the intensity map (Fig.~4a), leading to a large contribution to the joint density of states that provides the dominant contribution to $C(\textbf q,\omega)$ (Fig.~4b).  Although these high energy features are not as sharp as those in the low energy range, they are still well defined \cite{MCELROYAC}.  At high energies, similar behavior is seen in
the pseudogap phase (Figs.~4 c,d).  In that case, though, the dominance of antindoal sections in $C(\textbf q,\omega)$  starts to appear  at much lower energies ($\sim$ 25 meV) than in the
superconducting state since
the pseudogap does not remove as much spectral weight from the antinodal regions as the
superconducting gap does.

\begin{figure}
\includegraphics[width=3.4in]{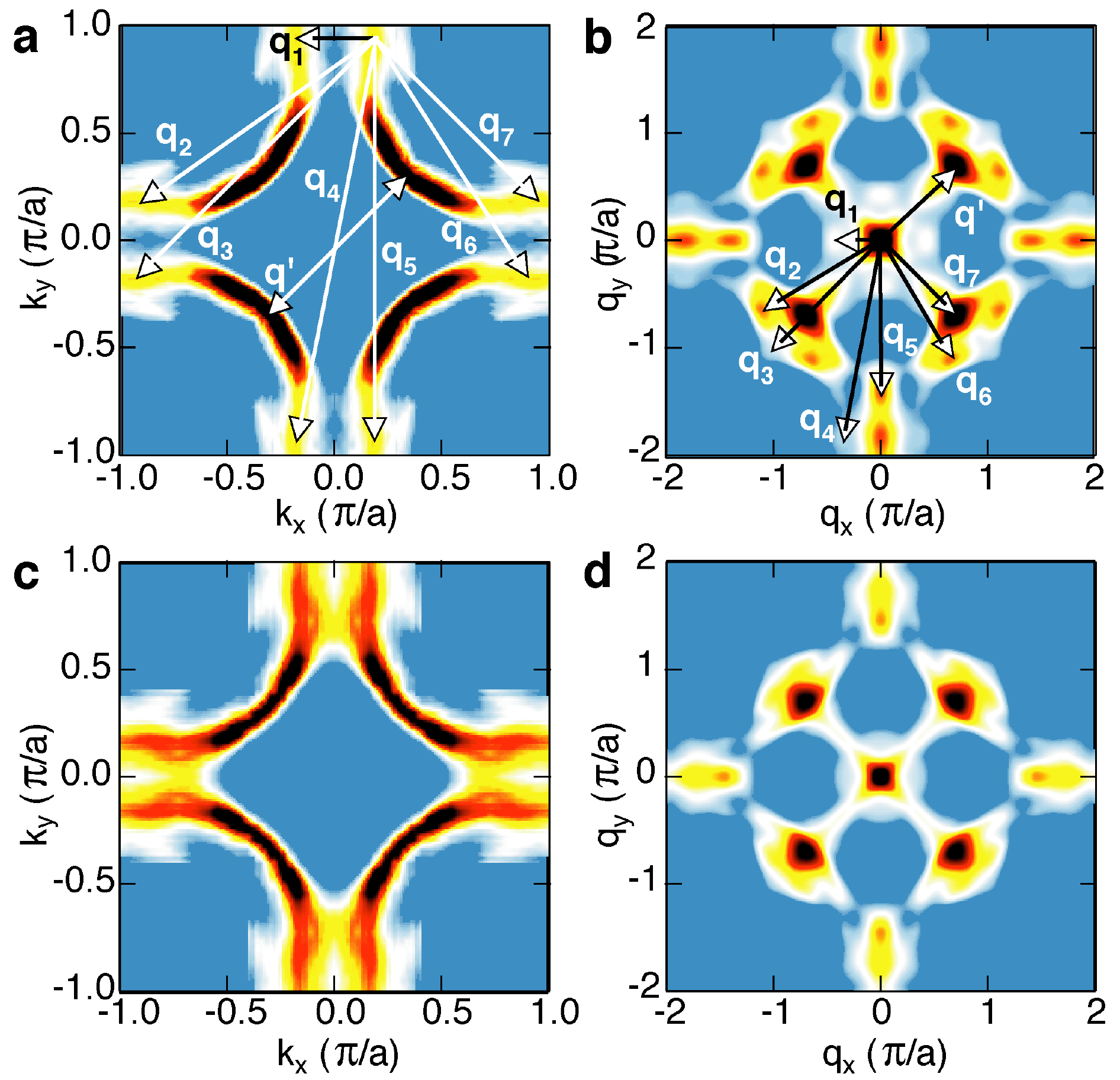}
\caption{(Color online) (a) Intensity map and (b) autocorrelation in the superconducting state at 90 meV.
(c,d) Analogous plots in the pseudogap phase at 90 meV.}   
\label{fig4.eps}
\end{figure}

In our previous work \cite{AC_ARPES}, we have shown that the ARPES autocorrelation analysis provides a model independent explanation for the origin of FT-STS peaks  at low energies, for both the 
superconducting state \cite{HOFFMAN,MCELROY} and for the pseudogap phase \cite{ALI}, based on a joint density of states perspective. A similar conclusion was reached in Ref.~\onlinecite{MCELROYAC}.  The autocorrelation analysis explains why the low energy FT-STS peaks are dispersive in the superconducting state and non-dispersive in the pseudogap phase. ARPES autocorrelations show all the features present in the FT-STS data in the superconducting state \cite{HOFFMAN,MCELROY}, along with $\textbf{q}^{\prime}$ which is not seen in FT-STS.  The lack of
observation of $\textbf{q}^{\prime}$ is a consequence of the c-axis tunneling matrix elements, which vanish
for nodal states \cite{ANDERSEN}.  The FT-STS data in the pseudogap phase \cite{ALI} does not show many of the features observed in the ARPES autocorrelations. But low temperature FT-STS data do  when the zero temperature pseuodgap regions of the sample \cite{STM_ZTPG} are analyzed.
In particular, the $\bf{q_1}$ peak in the autocorrelation along the bond direction (Figs. 2 and 4) is in quantitative agreement with the high energy FT-STS ($\textbf{q}=2\pi$/4.5) peak in these
regions. A careful look at the high energy FT-STS data of these regions \cite{STM_ZTPG} reveals that there is another peak (although very weak in intensity) at around $\textbf{q}=2\pi$/1.33 along the bond direction. This peak matches the $\bf{q_5}$ autocorrelation peak. However, there is an apparent contradiction. In the FT-STS data, the high energy features are not  visibly present in the superconducting regions of the sample; on the contrary, high energy ARPES autocorrelations show that  $\bf{q_1}$ and $\bf{q_5}$ are generic high energy features for both the pseudogap phase and the superconducting state. So we see that apart from some apparent discrepancies which remain to be understood, the ARPES autocorrelations can be directly connected to the FT-STS results. This agreement leads us to conclude that all features observed in the FT-STS data, both in the superconducting state and in the pseudogap phase \cite{HOFFMAN,MCELROY,ALI,STM_ZTPG}, are related to the $\textbf{k}$-dependent profiles of the joint density of states.

Interestingly, the FT-STS analysis of the zero temperature pseudogap regions (but {\it not} the superconducting regions) \cite{STM_ZTPG} also gives some
evidence for the anomalous `S' shaped dispersion arising from the interaction of electrons with collective modes (Fig.~3a).  We would suggest that future FT-STS studies look further into this issue.
In fact, recently, it has been observed that the FT-STS peaks in the high energy sector are visible in the superconducting regions if the inhomogeneity of the gap magnitude is taken into account in the analysis of the data \cite{LEE}.

In summary, given the strong correlation we find between the ARPES autocorrelation
and FT-STS data, we conclude that all of the results we find can be understood from a
joint density of states perspective. This had already been demonstrated at low energies in Refs.~\onlinecite{AC_ARPES,MCELROYAC}. Remarkably, we find that this correlation persists to higher energies, where many-body effects have a profound impact on the data. Our results support calculations that are based on a joint density of states interpretation of the FT-STS peaks \cite{BALATSKY}, in constrast to those proposals that invoke dynamic charge order \cite{RMP}.

We acknowledge useful discussions with J. C. Davis and A. Yazdani. This work was supported by NSF DMR-0606255, and the U.S. DOE, Office of Science, under Contracts No.~DE-AC02-06CH11357 (ANL) and W-7405-Eng-82 (Ames). The Synchrotron Radiation Center is supported by NSF  DMR-0537588.

\end{document}